\documentclass{article}
\usepackage[utf8]{inputenc}
\usepackage{algpseudocode}

\usepackage[final]{stylefiles/ACL2023}   

\usepackage{times}
\usepackage{latexsym}
\usepackage[utf8]{inputenc} 
\usepackage[T1]{fontenc}    
\usepackage{url}  
\usepackage{booktabs}       
\usepackage{amsfonts}       
\usepackage{amsmath}        
\usepackage{amsthm}         
\usepackage{amssymb}        
\usepackage{nicefrac}       
\usepackage{microtype}      
\usepackage{xcolor}         
\usepackage{colortbl}       
\usepackage{graphicx}       
\usepackage{subcaption}     
\usepackage{tabularx}
\usepackage{tabulary}
\usepackage{hyperref}
\usepackage{inconsolata}    
\usepackage{multirow}
\usepackage{listings}       
\usepackage{enumitem}       
\usepackage[normalem]{ulem} 
\usepackage{colortbl}       
\usepackage{makecell}
\usepackage{algorithm}
\usepackage{algpseudocode}
\usepackage{caption}
\usepackage{utfsym}

\usepackage[colorinlistoftodos,prependcaption,textsize=tiny]{todonotes}

\theoremstyle{definition}

\definecolor{backcolour}{rgb}{0.95,0.95,0.92}
\definecolor{arylideyellow}{rgb}{0.91, 0.84, 0.42}
\definecolor{cream}{rgb}{1.0, 0.99, 0.82}
\definecolor{darkblue}{rgb}{0, 0, 0.5}

\hypersetup{colorlinks=true, citecolor=darkblue, linkcolor=darkblue, urlcolor=darkblue}

\definecolor{codegreen}{rgb}{0,0.6,0}
\definecolor{codegray}{rgb}{0.5,0.5,0.5}
\definecolor{codepurple}{rgb}{0.58,0,0.82}
\definecolor{backcolour}{rgb}{0.95,0.95,0.92}

\lstdefinestyle{mystyle}{
    backgroundcolor=\color{backcolour},   
    commentstyle=\color{codegreen},
    keywordstyle=\color{magenta},
    numberstyle=\tiny\color{codegray},
    stringstyle=\color{codepurple},
    basicstyle=\ttfamily\footnotesize,
    breakatwhitespace=false,         
    breaklines=true,                 
    captionpos=b,                    
    keepspaces=true,                 
    numbers=left,                    
    numbersep=5pt,                  
    showspaces=false,                
    showstringspaces=false,
    showtabs=false,                  
    tabsize=2
}

\newcommand{\newcheckmark}{\usym{2713}}
\newcommand{\newcrossmark}{\usym{2717}}

\title{Leveraging LLM Reasoning Enhances Personalized Recommender Systems}

\author{
 \textbf{Alicia Y. Tsai*†\textsuperscript{1, 3}},
 \textbf{Adam Kraft*\textsuperscript{3}},
 \textbf{Long Jin\textsuperscript{2}},
 \textbf{Chenwei Cai\textsuperscript{2}},
\\
 \textbf{Anahita Hosseini\textsuperscript{3}},
 \textbf{Taibai Xu\textsuperscript{2}},
 \textbf{Zemin Zhang\textsuperscript{2}},
 \textbf{Lichan Hong\textsuperscript{3}},
\\
 \textbf{Ed H. Chi\textsuperscript{3}},
 \textbf{Xinyang Yi\textsuperscript{3}}
\\
\\
 \textsuperscript{1}University of California, Berkeley
 \textsuperscript{2}Google
 \textsuperscript{3}Google DeepMind
}

\begin{document}
\maketitle

\begingroup\def\thefootnote{*}\footnotetext{Equal contribution. Contact: aliciatsai@berkeley.edu, adamkraft@google.com.}
\begingroup\def\thefootnote{†}\footnotetext{Work done while at Google DeepMind.}\endgroup

\begin{abstract}
    Recent advancements have showcased the potential of Large Language Models (LLMs) in executing reasoning tasks, particularly facilitated by Chain-of-Thought (CoT) prompting. While tasks like arithmetic reasoning involve clear, definitive answers and logical chains of thought, the application of LLM reasoning in recommendation systems (RecSys) presents a distinct challenge. RecSys tasks revolve around subjectivity and personalized preferences, an under-explored domain in utilizing LLMs' reasoning capabilities. Our study explores several aspects to better understand reasoning for RecSys and demonstrate how task quality improves by utilizing LLM reasoning in both zero-shot and fine-tuning settings. Additionally, we propose \textbf{Rec-SAVER} (\textbf{Rec}ommender \textbf{S}ystems \textbf{A}utomatic \textbf{V}erification and \textbf{E}valuation of \textbf{R}easoning) to automatically assess the quality of LLM reasoning responses without the requirement of curated gold references or human raters. We show that our framework aligns with real human judgment on the coherence and faithfulness of reasoning responses. Overall, our work shows that incorporating reasoning into RecSys can improve personalized tasks, paving the way for further advancements in recommender system methodologies.
\end{abstract}

\section{Introduction}
\label{sec:intro}
The rapid advancement of Large Language Models (LLMs) has ushered in a new era of transformative capabilities, demonstrating the potential across a spectrum of applications. Recent progress has showcased their ability in reasoning tasks. Particularly, the advent of Chain-of-Thought (CoT) \cite{wei2022chain} and zero-shot CoT prompting \cite{kojima2022large} has provided a pathway for these models to engage in multi-step reasoning. Tasks examined in these studies, ranging from arithmetic reasoning \cite{amini-etal-2019-mathqa, cobbe2021training, hendrycks2021measuring} to commonsense question answering \cite{geva2021did, talmor-etal-2019-commonsenseqa}, typically demand clear, definitive answers and logical chains of thought. In contrast, the landscape of recommender systems (RecSys),  outlined in Figure \ref{fig:recsys_landscape}, introduces a nuanced challenge, where reasoning extends beyond objective criteria to encompass subjectivity and personalized user preferences. This aspect remains an  under-explored domain in leveraging the reasoning capabilities of LLMs. Prior works have utilized LLMs in recommender systems, employing techniques such as in-context learning and instruction tuning \cite{gao2023chat, geng2022recommendation, kang2023llms, zhang2023recommendation}. However, a comprehensive understanding of how LLMs execute reasoning in the context of personalized preferences remains elusive. Our work fills this gap by investigating how the reasoning capabilities of LLMs can enhance personalized recommendations in zero-shot and fine-tuning, resulting in improved task performance.

In contrast to the prediction task performance, objectively assessing the quality of reasoning presents challenges in the absence of curated gold standard references or human raters. To surmount this, we propose \textbf{Rec-SAVER}. This framework provides an efficient means of assessing the quality of LLM outputs, contributing to our understanding of LLM reasoning dynamics in personalized recommendation scenarios. We gauge human assessments on \textit{coherence}, \textit{faithfulness}, and \textit{insightfulness}. Our observations suggest that syntactic metrics such as BLEU and ROUGE, demonstrate suitability in evaluating the output faithfulness of LLMs. On the other hand, metrics like METEOR and BERTScore prove to be more adept at measuring coherence in the generated outputs. To the best of our knowledge, this is the first study that comprehensively examines the effects and quality of LLM reasoning for personalized RecSys tasks. In summary, our contributions are as follows:
\begin{enumerate}[topsep=0pt,itemsep=0ex,partopsep=1ex,parsep=0pt]
    \item We explore the utilization of LLMs for reasoning in personalized recommendations, showcasing notable task performance improvements in both zero-shot and fine-tuning scenarios.
    \item We demonstrate the effectiveness of using larger models to generate reasoning data, enhancing the performance and reasoning abilities of smaller fine-tuned models.
    \item We introduce \textbf{Rec-SAVER}, an automatic reasoning evaluation framework that doesn't require curated gold references. It offers meaningful insights into LLMs' reasoning capabilities, aligning with human judgment while providing cost and efficiency benefits.
\end{enumerate}
\section{Methodology}
\label{sec:method}

\begin{figure}[!h]
    \captionsetup{font=footnotesize}
    \centering
    \includegraphics[width=.48\textwidth]{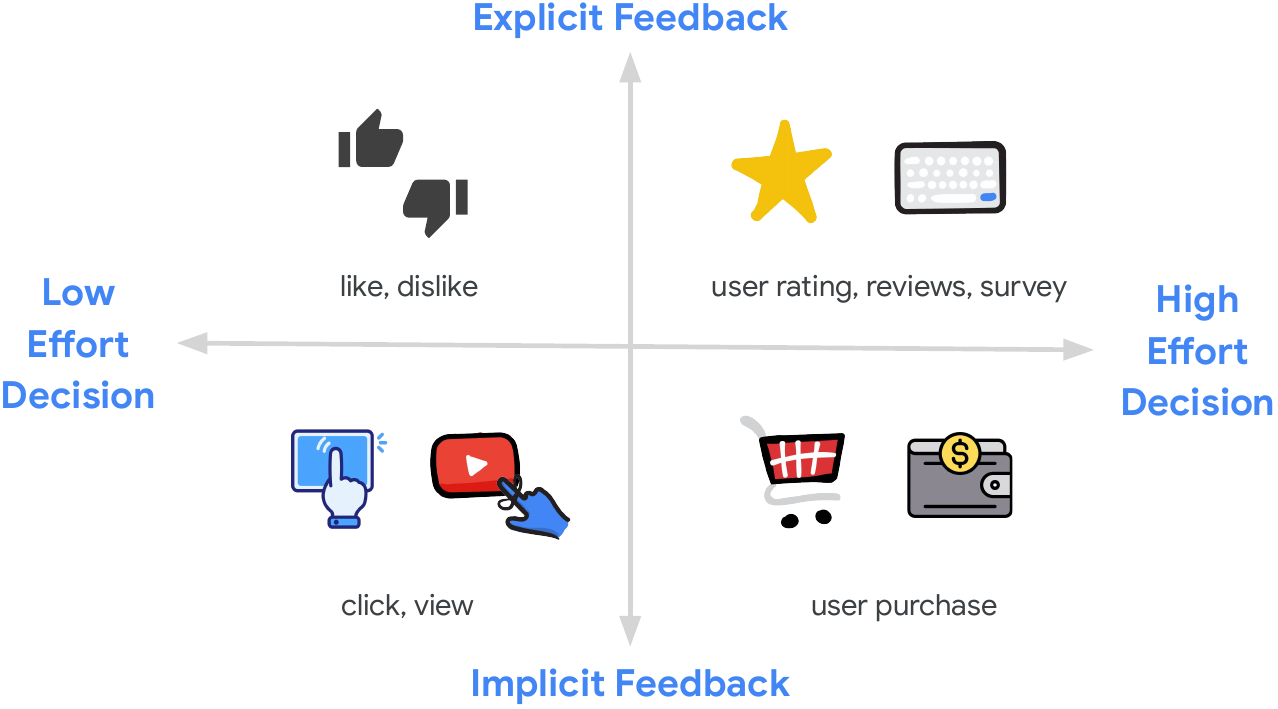}
    \caption{Landscape of recommender systems tasks, with user feedback extent on the vertical axis and decision-making effort on the horizontal. For example, a user clicking on websites requires low effort and does not provide much feedback about the user's satisfaction. Conversely, a user rating and reviewing products requires more effort and provides better satisfaction signals. }
    \label{fig:recsys_landscape}
\end{figure}

\subsection{Problem Setting}
The landscape of RecSys tasks, segmented along two axes, is illustrated in Figure \ref{fig:recsys_landscape}. In our experiments, we focus on the \textit{user rating prediction} task, which involves a high degree of both user decision making effort and collected user feedback, making this task well suited for exploring the extent of LLM reasoning for RecSys.

Let $\mathcal{R}$ represent a collection of user ratings and $\mathcal{D}$ denote a collection of user-written reviews for a set of items $\mathcal{I}$ belonging to a specific category (e.g., books), provided by users in $\mathcal{U}$. Each rating $\mathbf{r}_{u, i} \in \mathcal{R}$ is paired with a corresponding written review $\mathbf{d}_{u, i} \in \mathcal{D}$, reflecting the user's $u \in \mathcal{U}$ overall satisfaction with item $i \in \mathcal{I}$. Each item $i$ is associated with metadata $\mathcal{M}_i$, comprising details such as title, description, category, brand, and price. The user's purchase history $\mathcal{H}_u = (h_{u, 1}, h_{u, 2}, \dots, h_{u, t})$ constitutes a chronologically ordered collection of past purchases. Each past purchase $h_{u, j} = (\mathcal{M}_j, \mathbf{r}_{u, j}, \mathbf{d}_{u, j})$ represents a triplet comprising the metadata for the purchased item $j$, the user's rating for the purchase item, and the user's review for that item. The primary objective of the rating prediction task is to forecast the unknown ratings for items that users have not yet reviewed. The objective can be formalized as follows:
\begin{equation}
\begin{aligned}
\label{eq:rating-pred}
    \hat{\mathbf{r}}_{u, i} = \arg\max_k \mathbb{P}(\mathbf{r}_{u, i} = k \mid \mathcal{H}_u, \mathcal{M}_i), \\
    \text{where} \;\; i \notin \mathcal{H}_u, \; k \in \{1, 2, 3, 4, 5\}.
\end{aligned}
\end{equation}
In this equation, $\hat{\mathbf{r}}_{u, i}$ represents the predicted rating for user $u$ on item $i$, chosen from the set of possible ratings 1 to 5. This prediction is based on the user's purchase history $\mathcal{H}_u$ and the item's metadata $\mathcal{M}_i$, where the item $i$ has not been previously reviewed by user $u$. Recent advancements in recommendation systems utilize LLMs to model the rating prediction task described by equation \eqref{eq:rating-pred}, denoted as $\hat{\mathbf{r}}_{u, i} = \text{LLM}(\mathcal{H}_u, \mathcal{M}_i)$.

\subsection{Zero-shot Learning with Reasoning} 
\begin{figure}[!htbp]
    \captionsetup{font=footnotesize}
    \centering
    \includegraphics[width=.45\textwidth]{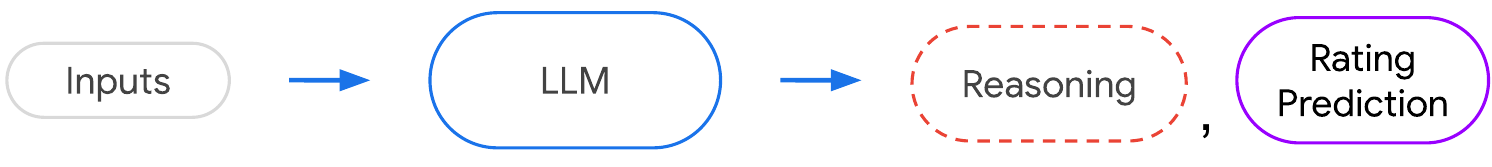}
    \caption{We prompt the LLM to generate a reasoning output prior to outputting the final task prediction.}
    \label{fig:reasoning_via_prompting}
\end{figure}
\label{sec:zero-shot}

As shown in Figure \ref{fig:reasoning_via_prompting}, we employ zero-shot CoT prompting strategies \cite{kojima2022large} to guide the LLM in generating a reasoning response, denoted as $\hat{\mathbf{s}}_{u, i}$, alongside a rating prediction $\hat{\mathbf{r}}_{u, i}$ for a given user $u$ and recommended item $i$: 
\begin{align}
\bigl \langle \hat{\mathbf{s}}_{u, i}, \; \hat{\mathbf{r}}_{u, i} \bigr \rangle = \text{LLM}(\mathcal{H}_u, \mathcal{M}_i).
\end{align}
Our prompt consists of four key elements: a \textit{preamble}, the \textit{user history} $\mathcal{H}_u$, the \textit{new item metadata} $\mathcal{M}_i$, and a \textit{task description}. The \textit{preamble} provides context for the subsequent information and establishes the rating scale ranging from 1 to 5. Following the \textit{preamble}, the \textit{user history} $\mathcal{H}_u$ is presented sequentially, detailing the user's past interactions. A new item $i$ is then introduced along with it's metadata $\mathcal{M}_i$ before the \textit{task description}, prompting the model to make predictions. The \textit{task description} also delineates the output requirements for the model responses. An abstract prompt template is illustrated in Table \ref{tab:reason-gen-prompt-abstract}, while more prompt details are provided in Appendix \ref{app:reasoning-generation-prompt}. Unlike traditional RecSys modeling techniques, our approach leverages natural language presentation for all information. This enables a more intuitive representation of rich content as natural language, as opposed to numerical IDs, enabling a more encompassing understanding of information \cite{geng2022recommendation}.

\begin{table}
    \centering
    \footnotesize
    \captionsetup{font=footnotesize}
    \caption{\label{tab:reason-gen-prompt-abstract} Abstract prompt template guiding our zero-shot approach, prompting the model to reason by leveraging user history and inferring preferences before making predictions.}
    \resizebox{0.48\textwidth}{!}{%
    \begin{tabulary}{0.64\textwidth}{lL}
    \toprule
    \textit{Preamble} & \textit{e.g.} Here is information about a user and a new product ... \\
    \textit{User History} & $h_{u, 1} = (\mathcal{M}_1, \mathbf{r}_{u, 1},  \mathbf{d}_{u, 1}), \dots, h_{u, t} = (\mathcal{M}_t,  \mathbf{r}_{u, t},  \mathbf{d}_{u, t})$ \\
    \textit{New Item} & $\mathcal{M}_i$, \textit{e.g.} title, brand, category, ... \\
    \textit{Task Description} & \textit{e.g.} Given the user's past purchase history [...] \\ & how they will rate the new item? [...] \\ & After your reasoning, predict a numerical rating. \\
    \bottomrule
    \end{tabulary}}
\end{table}

\subsection{Fine-tuning with Reasoning}
\label{sec:fine-tune-reason}

\begin{figure}[!h]
    \centering
    \captionsetup{font=footnotesize}
    \includegraphics[width=.48\textwidth]{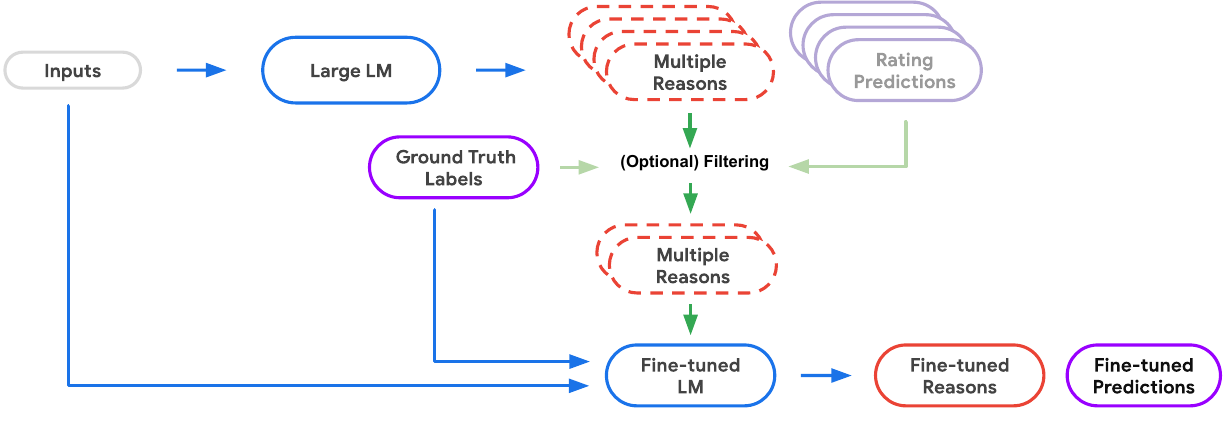}
    \caption{Fine-tuning a model with reasoning. We first collect multiple reasoning samples by prompting a Large LM. We then use the reasoning samples combined with the original rating ground truth labels to fine-tune a different (potentially smaller) LM. We can optionally filter the reasoning outputs by comparing the Large LM rating predictions with the ground truth ratings.}
    \label{fig:fine-tune-overview}
\end{figure}

Zero-shot learning with CoT prompting can be computationally intensive. Hence, fine-tuning with domain-specific datasets has emerged as a pragmatic strategy, especially when leveraging smaller pre-trained models \cite{flan-t5}. Our interest lies in investigating whether training with reasoning outputs can further enhance task performance. Building on the prompting methods outlined in Section \ref{sec:zero-shot}, we collect reasoning outputs generated by a larger language model to serve as training data for fine-tuning smaller models. For each user-item pair $(u, i)$ with input $(\mathcal{H}_u, \mathcal{M}_i)$, we gather multiple reasoning responses and rating predictions by adjusting a decoding temperature parameter $T > 0$ during generation \cite{holtzman2019curious}:
\begin{align}
    \label{eq:reason-gen}
    \bigl \langle \hat{\mathbf{s}}_{u, i}^{m}, \; \hat{\mathbf{r}}_{u, i}^{m} \bigr \rangle = \text{LLM}(\mathcal{H}_u, \mathcal{M}_i),
\end{align}
where $m = 1, 2, \dots, M$ indexes the $M$ candidate output pairs sampled from the decoder. This process yields a diverse set of reasoning paths, which is particularly advantageous for personalized recommendations, recognizing that the same rating can stem from various personal preferences and reasons. Optionally, we can use different methods to filter out reasoning responses $\hat{\mathbf{r}}_{u,i}^{m}$ that do not align with the ground truth $\mathbf{r}_{u, i}$. The fine-tuned model is then trained using the reasoning responses $\hat{\mathbf{s}}_{u, i}^{m}$ and the real ground truth rating label $\mathbf{r}_{u,i}$ as targets. The overall method is illustrated in Figure \ref{fig:fine-tune-overview}.

\section{Rec-SAVER: Evaluation of Reasoning}
\label{sec:unsupervised-reason-eval}

\begin{figure*}[!h]
    \centering
    \captionsetup{font=footnotesize}
    \includegraphics[width=0.85\textwidth]{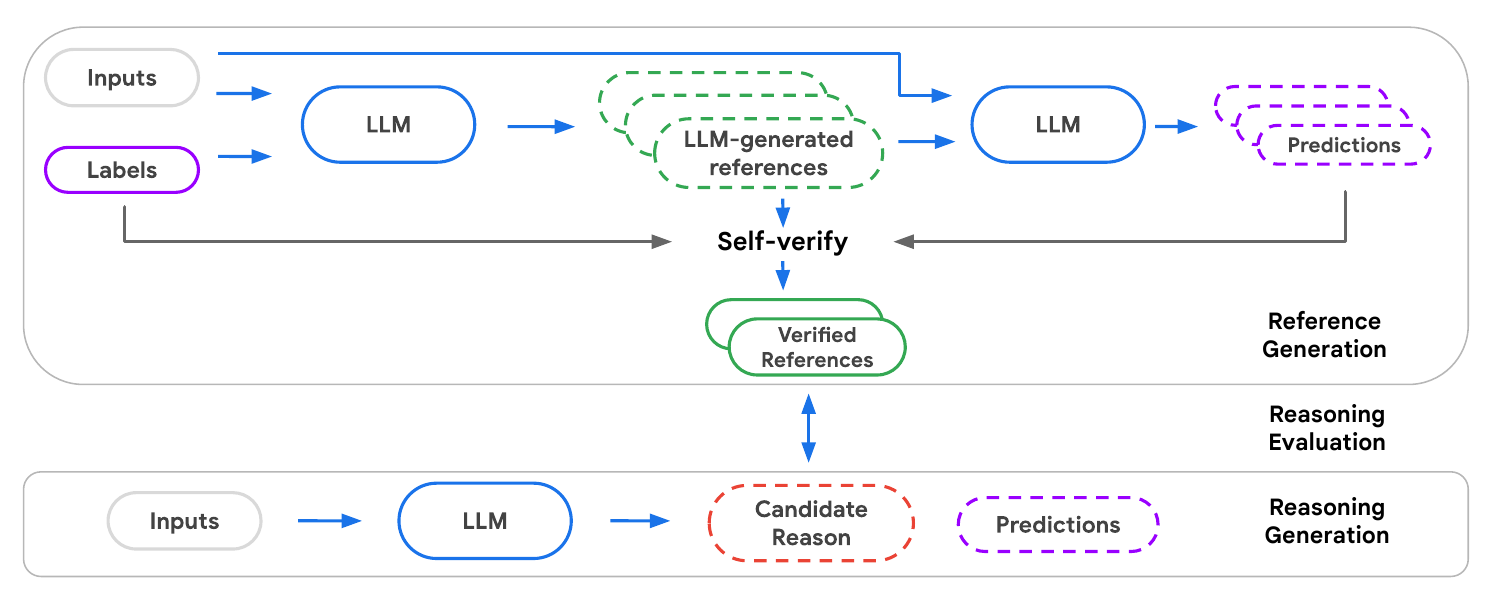}
    \caption{Overview of \textbf{Rec-SAVER} utilizing LLM-generated references and LLM self-verification. The first LLM call uses the ground truth rating labels as additional input to generate post hoc reasoning generated reference. We then do a subsequent LLM call passing in the generated reasoning reference and collect a new rating prediction. We keep only the predictions where the final rating prediction matches the ground truth rating label as our verified references. These verified references are then used to evaluate the reasoning outputs from other LLMs.}
    \label{fig:reason_eval}
\end{figure*}

In contrast to reasoning processes for solving mathematical problems or general question answering tasks, reasoning in RecSys rating prediction is highly subjective and personalized for individual users. Unlike in other domains where humans can provide reasoning steps and verify their validity, resulting in curated gold references, such references are challenging to obtain in RecSys due to the subjective nature of user preferences. To address this challenge, we propose a system called \textbf{Rec-SAVER}: \textbf{Rec}ommender \textbf{S}ystems \textbf{A}utomatic \textbf{V}erification and \textbf{E}valuation of \textbf{R}easoning. Rec-SAVER aims to automatically generate good reasoning references specifically tailored for RecSys tasks. These references can then be utilized to quantitatively evaluate the quality of reasoning responses generated by LLMs. Additionally, we conduct a human study to demonstrate the alignment of our method with real human judgment, thus providing validation for the effectiveness of Rec-SAVER in assessing reasoning quality in RecSys applications.

\subsection{Reference Generation with Self-Verification}
The core concept of Rec-SAVER involves a two-step process leveraging LLM-generated explanations and LLM self-verification. As illustrated in Figure \ref{fig:reason_eval}, we present the LLM with a user-item pair $(\mathcal{H}_u, \mathcal{M}_i)$. Additionally the target user rating $\mathbf{r}_{u, i}$ is also provided as input. The model is instructed to provide a post hoc explanation, describing why the user assigned such a rating based on the given user history and new item information. We denote this post hoc explanation generated by LLM as $\hat{\mathbf{g}}_{u, i}$. Note that this is different from the aforementioned reasoning $\hat{\mathbf{s}}_{u, i}$, where the ground truth rating $\mathbf{r}_{u, i}$ is not included as an input. Similar to the approach in Section~\ref{sec:fine-tune-reason} where multiple responses are sampled, we generate $N$ different explanations $\hat{\mathbf{g}}_{u, i}^{n}$ where $n = 1, 2, \dots, N$. These post hoc explanations are then passed onto a verification process.

To ensure the credibility and consistency of these LLM-generated explanations $\hat{\mathbf{g}}_{u, i}^{n}$, we implement a self-verification step atop the previous explanation generation process. The self-verification step involves making a second call to the same LLM, inputting the user-item information $(\mathcal{H}_u, \mathcal{M}_i)$, and the explanations $\hat{\mathbf{g}}^n_{u, i}$ generated from the previous call. The model is then tasked with making a rating prediction based on the user history, new item information, and the post hoc explanation, formally defined as $\tilde{\mathbf{r}}^n_{u, i} = \text{LLM}(\mathcal{H}_u, \mathcal{M}_i, \hat{\mathbf{g}}^n_{u, i})$. However, in practice, we have observed that many explanations $\hat{\mathbf{g}}^n_{u, i}$ contain text snippets such as ``\textit{the user gave a rating of 5 because ...}'', which can lead to information leakage. To prevent $\hat{\mathbf{g}}^n_{u, i}$ from directly leaking the ground truth, we employ a simple post-processing step by removing sentences that mention ``\textit{a rating of}'', ``\textit{stars}'', or ``\textit{scores}'' before performing the prediction. In future work we aim to improve upon this manual process to fully ensure the removal of information leakage.

We then validate whether the new rating $\tilde{\mathbf{r}}^n_{u, i}$ matches the original ground truth $\mathbf{r}_{u, i}$. Explanations $\hat{\mathbf{g}}^n_{u, i}$ that pass the self-verification step are retained as the final verified references, constituting a diverse pool of LLM-generated references $\hat{\mathcal{G}}$. This two-step process follows the intuition that good explanations based on the given information and the ground truth should enable the model to make a correct prediction. By validating the predictions based on the generated explanations against the ground truth ratings, we ensure that only high-quality explanations are retained to serve as the final references. These verified references then serve as proxies for the unknown set of gold references $\mathcal{G}$. Since the self-verification may result in different verified references per sample, we may have varying numbers of final references per sample. The full reference generation process is summarized in Algorithm \ref{alg:ref-gen}.

\begin{algorithm}[!t]
\caption{Reference generation with self-verification}
\label{alg:ref-gen}
\begin{algorithmic}[1]
\State \textbf{Inputs:} $N$
\State $\hat{\mathcal{G}} \gets \emptyset$  \Comment verified references
\For{$(\mathcal{H}_u, M_{i}, \mathbf{r}_{u, i})$ in dataset}
\For{$n = 1 \dots N$}
    \State $\hat{\mathbf{g}}^n_{u, i} \gets \text{LLM}(\mathcal{H}_u,
    \mathcal{M}_i, \mathbf{r}_{u, i})$
    \State $\hat{\mathbf{g}}^n_{u, i} \gets \text{post-process}(\hat{\mathbf{g}}^n_{u, i})$
    \State $\Tilde{\mathbf{r}}^n_{u, i} \gets \text{LLM}(\mathcal{H}_u, \mathcal{M}_i, \hat{\mathbf{g}}^n_{u, i})$
    \If{$\Tilde{\mathbf{r}}^n_{u, i} = \mathbf{r}_{u, i}$}
        \State $\hat{\mathcal{G}} \gets \hat{\mathcal{G}} \cup \{\hat{\mathbf{g}}^n_{u, i}\}$
    \EndIf
\EndFor
\EndFor
\end{algorithmic}
\end{algorithm}

\subsection{Human Judgment Alignment Study}
\label{sec:human-judgment-study}
To gauge the effectiveness of Rec-SAVER, we conduct a study to evaluate how our proposed method aligns with human judgment regarding the candidate reasons generated by LLMs. This study aims to provide insight into the reliability and validity of our proposed method. During the study, human raters are presented with sample input prompts and the reasoning outputs $\hat{\mathbf{s}}_{u, i}$ generated by LLMs. It is important to note that these reasoning outputs $\hat{\mathbf{s}}_{u, i}$ are produced using only user history $\mathcal{H}_u$ and item metadata $\mathcal{M}_i$ as inputs; no ground truth rating $\mathbf{r}_{u, i}$ is provided to the LLM. No ratings (ground truth or LLM predicted) are shown to the human raters. Human raters are asked to assess the reasoning outputs based on the following dimensions:
\begin{itemize}[topsep=0pt,itemsep=0ex,partopsep=1ex,parsep=0pt]
    \item \textbf{Coherence (5-point Likert)}: Evaluate whether the generated reasoning makes sense and follows a clear and coherent logical flow that reflects the reasons behind the user’s preference.
    
    \item \textbf{Faithfulness (Binary)}: Examine the presence of hallucination in the generated reasoning and whether it contains fabricated information.
    
    \item \textbf{Insightfulness (5-point Likert)}: Assess the degree to which the generated reasoning delivers valuable, informative, interesting or delightful insights into the user’s preferences and purchasing patterns.
\end{itemize}
\section{Experiments}
\label{sec:experiments}
\subsection{Data Preparation and Tasks Setup}

Experiments are conducted on the Amazon product review dataset\footnote{\url{https://cseweb.ucsd.edu/~jmcauley/datasets/amazon\_v2/}}, which is a widely recognized benchmark in RecSys. This dataset offers comprehensive user feedback, including ratings and review text, as well as detailed product metadata such as descriptions, category information, price, and brand \cite{ni2019justifying}. We focus our experiments on the rating prediction task, conducting evaluations in two distinct domains: \textsc{Beauty} and \textsc{Movies/TV}. To understand the extent that LLMs can understand user preferences, we filter examples where $4 \le |\mathcal{H}_u| \le 10$. The lower threshold ensures we have enough past purchases to see trends and patterns while the higher threshold prevents inputs from exceeding the LLM context window. The original label distribution is heavily skewed towards positive ratings, with a rating of 5 accounting for over 60 \% of the data. We perform random subsampling to create a fully balanced dataset with an even label distribution, resulting in 4,000 training examples (800 per label) and 500 test examples (100 per label). The training split is used to test out different prompts for zero-shot learning and as training examples for the fine-tuning experiments. The user sets are mutually exclusive between training and test. This setup allows us to better test and understand the capabilities of LLM reasoning, while we acknowledge this may not reflect a full real world scenario.

In following sections, we report the rating task prediction metrics, including multi-class and binary metrics. Specifically, we compare the model's output $\hat{\mathbf{r}}_{u, i}$ against the ground truth rating $\mathbf{r}_{u, i}$. The binary metrics were calculated using a cutoff threshold of $\hat{\mathbf{r}}_{u, i} > 3$. The quality of the reasoning outputs will be analyzed in more detail later on in Section \ref{sec:reasoning-eval}.

\begin{table}[!h]
    \centering
    \captionsetup{font=footnotesize}
    \scriptsize
    \caption{\label{tab:task-description-example} Example task description prompt. The highlighted text represents portions removed when prompting the model to make predictions without the intermediate reasoning step.}
    \begin{tabulary}{.48\textwidth}{L}
        \toprule
        \textbf{Task Description} \\
        \midrule
        Given the user's past purchase history [...] how they will rate the new item? [...] 
        \colorbox{yellow!50}{After your reasoning,} predict a numerical rating.
        \newline
        === Please follow the format below: ===
        \newline
        \colorbox{yellow!50}{\#\#\# Reason \#\#\#}
        \newline
        \colorbox{yellow!50}{Write your reasoning explanation here.}
        \newline
        \#\#\# Rating \#\#\# \newline
        Give a single numerical rating, e.g. 1 \\
        \bottomrule
    \end{tabulary}
\end{table}

\begin{table*}[t]
    \centering
    \captionsetup{font=footnotesize}
    \footnotesize
    \renewcommand{\arraystretch}{0.90}
    \setlength{\tabcolsep}{5pt}
    \caption{\label{tab:zero-shot-ablation-all} Comparisons and ablation studies on zero-shot learning with PaLM 2-M, investigating the role of reasoning outputs and input features. ``Our Method'' denotes zero-shot chain-of-thought with reasoning output, while ``No Reasoning Outputs'' refers to the rating prediction-only task. Additionally, we examine the impact of removing user reviews, user ratings, and item descriptions from the input to the LLM. The ``Naive Baseline'' uses the historical rating average of the user as the prediction. We also include results for one-shot learning.}
    \resizebox{.9\textwidth}{!}{%
    \begin{tabulary}{1.2\textwidth}{LLCCCCCCCCC}
        \toprule
        & \thead{\textbf{Method}} & 
        \thead{\textbf{Binary} \\ \textbf{Acc.}} &
        \thead{\textbf{Binary} \\ \textbf{F1}} &
        \thead{\textbf{Multi.} \\ \textbf{Acc.}} & 
        \thead{\textbf{Multi.} \\ \textbf{MAE $\downarrow$}} &
        \thead{\textbf{Multi.} \\ \textbf{RMSE $\downarrow$}} &
        \thead{\textbf{ROUGE-1} \\ \textbf{F1}} & 
        \thead{\textbf{METEOR}} &
        \thead{\textbf{BLEU}} &
        \thead{\textbf{BERT} \\ \textbf{Score}} \\
        \midrule
        \multirow{7}{*}{\rotatebox[origin=c]{90}{\textsc{Beauty}}}
        & Naive Baseline (Avg.) & 0.52 & 0.60 & 0.25 & 1.35 & 1.75 & - & - & - & - \\
        \cmidrule(lr){2-11}
        & Our Method (zero-shot CoT) & \textbf{0.56} & \textbf{0.62} & \textbf{0.37} & \textbf{1.14} & \textbf{1.60} & 0.236 & 0.503 & 0.339 &	0.665\\
        & - No Reasoning Outputs & 0.49 & 0.57 & 0.23 & 1.35 & 1.70 & - & - & - & -  \\
        & - No Review & 0.48 & 0.57 & 0.21 & 1.35 & 1.69 & \textbf{0.237} & \textbf{0.507} & 0.337 & \textbf{0.667} \\
        & - No Review, No Rating &  0.43 & 0.53 & 0.19 & 1.42 & 1.75 & 0.215 & 0.494 & 0.331 & 	0.660 \\
        & - No Item Description & 0.48 & 0.57 & 0.21 & 1.33 & 1.66 & 0.235 & 0.504 &  \textbf{0.340} & \textbf{0.667} \\
        & One-shot & 0.43 & 0.57 & 0.26 & 1.52 & 1.97 & 0.225 & 0.502 & 0.335 & 0.664 \\
        \midrule
        \multirow{7}{*}{\rotatebox[origin=c]{90}{\textsc{Movies/TV}}}
        & Naive Baseline (Avg.) & 0.59 & 0.63 & 0.30 & 1.21 & 1.63 & - & - & - & - \\
        \cmidrule(lr){2-11}
         & Our Method (zero-shot CoT) & \textbf{0.62} & \textbf{0.66} & \textbf{0.40} & \textbf{1.06} & \textbf{1.53} & \textbf{0.194} & \textbf{0.465} & \textbf{0.296} & \textbf{0.647} \\
        & - No Reasoning Output &  0.59 & 0.63 & 0.29 & 1.18 & 1.56 & - & - & - & - \\
        & - No Review & 0.58 & 0.63 & 0.28 & 1.20 & 1.58 & 0.173 & 0.452 & 0.291 & 0.641 \\
        & - No Review, No Rating & 0.43 & 0.54 & 0.20 & 1.42 & 1.75 & 0.150 & 0.434 & 0.283 & 0.633 \\
        & - No Item Description & 0.54 & 0.62 & 0.28 & 1.22 & 1.60 & 0.183 & 0.460 & \textbf{0.296} & \textbf{0.647} \\
        & One-shot & 0.47 & 0.59 & 0.23 & 1.32 & 1.68 & 0.182 & 0.452 & 0.276 & 0.641 \\
        \bottomrule
    \end{tabulary}}
\end{table*}

\subsection{Zero-shot Learning Improves with Reasoning}
In the following zero-shot experiments, we utilize the PaLM 2-M LLM \cite{anil2023palm}, a highly capable model trained on a broad set of languages and tasks, including reasoning tasks. Initially, we investigate the impact of prompting the model to engage in reasoning prior to prediction (zero-shot chain-of-thought) compared to direct prediction (zero-shot). Table \ref{tab:task-description-example} shows the differences between the final task description of the input prompt for zero-shot CoT and direct prediction.  Subsequently, Table \ref{tab:zero-shot-ablation-all} outlines the outcomes of different zero-shot ablation studies. We also compare against a naive baseline, where we use the historical rating average of the user's history as a prediction for the future item. We observe a notable performance improvement across both product domains when the model is guided to output reasoning alongside the prediction (``Our Method (zero-shot CoT)'' vs. ``No Reasoning Outputs''). This suggests that personalized tasks are inherently difficult for LLMs to solve without further guidance such as engaging in an intermediate reasoning step.

\vspace{-.2cm}
\paragraph{Impact of explicit user feedback.}
A user's past purchase, denoted as $h_t = (\mathcal{M}_j, \mathbf{r}_{u, j}, \mathbf{d}_{u, j}) \in \mathcal{H}_u$, encapsulates the user-item relations, providing explicit user feedback. To understand the helpfulness of this feedback, we conduct ablation studies. In the first case, we eliminate the written reviews $\mathbf{d}_{u, j}$ from the purchase history where $h_t = (\mathcal{M}_j, \mathbf{r}_{u, j})$. In the second case, we further eliminate the numerical ratings $\mathbf{r}_{u, j}$ from $h_t$, resulting in  $h_t = (\mathcal{M}_j)$. The first case mirrors a common scenario in RecSys where only numerical rating information is available, while the second case simulates scenarios where only implicit feedback from a user, in the form of past purchases, is accessible. Table \ref{tab:zero-shot-ablation-all} presents the ablation results, highlighting a significant performance drop when the review text is excluded from $h_t$ (``No Review''). The performance declines further when both reviews and ratings are excluded (``No Review, No Rating''). When only the written review text is removed, the results are similar to direct predictions made without reasoning (``No Review'' vs ``No Reasoning Outputs) and sometimes even worse than naive average baseline. This indicates that review text is essential for utilizing the reasoning capabilities of LLMs. Without user reviews, the model lacks detailed insights into past user interactions and can only rely on numerical ratings, resulting in performance similar to or worse than ``No Reasoning Outputs'' and the naive average baseline.

Furthermore, when both reviews and ratings are unavailable, the outcomes are akin to random guessing, as evidenced by the multi-class accuracy hovering around 0.2. This performance is strictly worse than the naive baseline and direct prediction without the reasoning outputs. Our ablation studies suggest that while the LLM can estimate some user preference information simply from the numerical user rating, it benefits even more when we have the full written user reviews and a guided reasoning step. Review texts help discern nuanced details about a user's specific preferences. For instance, a user might rate the movie ``\textit{Top Gun}'' as 5 stars for various reasons, such as their interest in airplanes, admiration for Tom Cruise, or a fondness for action movies in general. These preferences may be explicitly stated in the review text, enabling the LLM to make more informed decisions.

\vspace{-.2cm}
\paragraph{Effect of Pre-trained Knowledge.}
In addition to analyzing the impact of excluding explicit user feedback, we also investigate the removal of item descriptions. We found that while performance decreases in both domains when item descriptions are unavailable, the decline is less pronounced for \textsc{Movies/TV} compared to \textsc{Beauty}. This finding suggests that the LLM (PaLM 2-M) possesses a more extensive knowledge base in \textsc{Movies/TV} domain, enabling it to infer information about the recommended movies or TV shows even in the absence of descriptions.

\vspace{-.2cm}
\paragraph{One-shot learning.}
We compare one-shot learning against our zero-shot results (Table \ref{tab:zero-shot-ablation-all}). Surprisingly, the one-shot results are significantly worse, yielding only slight improvements over random performance. We believe the inclusion of a single example considerably increases the input text length. This could potentially hinder the model's ability to disentangle information from the provided example and the actual inputs $(\mathcal{M}_i, \mathcal{H}_u)$, resulting in poorer performance.

\begin{table*}[!h]
    \centering
    \captionsetup{font=footnotesize}
    \footnotesize
    \renewcommand{\arraystretch}{0.95}
    \setlength{\tabcolsep}{5pt}
    \caption{\label{tab:fine-tune-model-size} Fine-tuning results on Flan-T5, comparing different model sizes and fine-tuning with and without reasoning (predict numerical rating only). We also show the XL model results without any fine-tuning. Without fine-tuning, the model was unable to follow instructions with reasoning, and therefore we only show results without reasoning.}
    \resizebox{1\textwidth}{!}{%
    \begin{tabulary}{1.3\textwidth}{LLLLCCCCCCCCCC}
        \toprule
        & \thead{\textbf{Model} \\ \textbf{Size}} & 
        \thead{\textbf{Reas-} \\ \textbf{oning}} &
        \thead{\textbf{Binary} \\ \textbf{Acc.}} &
        \thead{\textbf{Binary} \\ \textbf{F1}} &
        \thead{\textbf{Binary} \\ \textbf{AUC}} &
        \thead{\textbf{Multi.} \\ \textbf{Acc.}} & 
        \thead{\textbf{Multi.} \\ \textbf{AUC}} &
        \thead{\textbf{Multi.} \\ \textbf{MAE $\downarrow$}} &
        \thead{\textbf{Multi.} \\ \textbf{RMSE $\downarrow$}} &
        \thead{\textbf{BLEU}} &
        \thead{\textbf{ROUGE-1} \\ \textbf{F1}} & 
        \thead{\textbf{METEOR}} &
        \thead{\textbf{BERT} \\ \textbf{Score}} \\
        \midrule
        \multirow{5}{*}{\rotatebox[origin=c]{90}{\textsc{Beauty}}}
         & Small & \newcheckmark & 0.62 & 0.53 & 0.65 & 0.30 & 0.63 & 1.35 & 1.84 & 0.225 & 0.499 & 0.342 & 0.663 \\
          & Base & \newcheckmark & 0.59 & 0.47 & 0.66 & 0.27 & 0.64 & 1.37 & 1.83 & 0.239 & 0.507 & \textbf{0.344} & \textbf{0.667} \\
          & Large & \newcheckmark & 0.64 & 0.59 & 0.67 & \textbf{0.33} & 0.65 & 1.26 & 1.73 & 0.240 &  0.506 & 0.343 & 0.666 \\
          & XL & \newcheckmark & \textbf{0.67} & \textbf{0.61} & \textbf{0.78} & 0.30 & \textbf{0.69} & \textbf{1.24} & \textbf{1.68} & \textbf{0.241} &  \textbf{0.510} & 0.339 & \textbf{0.667} \\
        & XL & \newcrossmark & 0.55 & \textbf{0.61} & 0.74 & 0.28 & 0.67 & 1.31 & 1.75 & - & - & - & - \\
        \cmidrule(lr){2-14}
        & XL (no fine-tuning) & \newcrossmark & 0.55 & 0.40 & 0.56 & 0.22 & 0.56 & 1.64 & 2.09 & - & - & - & - \\
        \midrule
        \multirow{5}{*}{\rotatebox[origin=c]{90}{\textsc{Movies/TV}}}
          & Small & \newcheckmark  & 0.60 & 0.55 & 0.70 & 0.33 & 0.66 & 1.23 & 1.71 & 0.137 & 0.423 & 0.272 & 0.627 \\
          & Base & \newcheckmark & \textbf{0.65} & 0.59 & \textbf{0.72} & \textbf{0.34} & \textbf{0.68} & 1.18 & 1.65 & 0.153 & 0.438 & 0.279 & 0.634 \\
          & Large & \newcheckmark & 0.64 & 0.58 & \textbf{0.72} & 0.32 & 0.67 & 1.23 & 1.70 & \textbf{0.165} & 0.448 & \textbf{0.286} & 0.639 \\
          & XL & \newcheckmark & \textbf{0.65} & \textbf{0.61} & \textbf{0.72} & \textbf{0.34} & 0.67 & \textbf{1.17} & \textbf{1.64} & \textbf{0.165} &  \textbf{0.449} & \textbf{0.286} & \textbf{0.643} \\
          & XL & \newcrossmark & 0.62 & 0.57 & 0.70 & 0.33 & 0.66 & 1.27 & 1.75 & - & - & - & - \\
          \cmidrule(lr){2-14}
        & XL (no fine-tuning) & \newcrossmark & 0.61 & 0.43 & 0.61 & 0.23 & 0.61 & 1.56 & 2.00 & - & - & - & - \\
        \bottomrule
    \end{tabulary}}
\end{table*}

\begin{table*}[!h]
    \centering
    \captionsetup{font=footnotesize}
    \footnotesize
    \renewcommand{\arraystretch}{0.95}
    \setlength{\tabcolsep}{5pt}
    \caption{\label{tab:fine-tune-more-samples-and-filtering} Comparison of fine-tuning Flan-T5 XL model with multiple reasoning paths per user-item pair and with different filtering methods. PaLM 2-M zero-shot (no fine-tuning) results are included for comparison.}
    \resizebox{\textwidth}{!}{%
    \begin{tabulary}{1.2\textwidth}{LCLCCCCCCCCCCC}
        \toprule
        & \thead{\textbf{Samples}} & 
        \thead{\textbf{Filter}} &
        \thead{\textbf{Binary} \\ \textbf{Acc.}} &
        \thead{\textbf{Binary} \\ \textbf{F1}} &
        \thead{\textbf{Binary} \\ \textbf{AUC}} &
        \thead{\textbf{Multi.} \\ \textbf{Acc.}} & 
        \thead{\textbf{Multi.} \\ \textbf{AUC}} &   
        \thead{\textbf{Multi.} \\ \textbf{MAE $\downarrow$}} &
        \thead{\textbf{Multi.} \\ \textbf{RMSE $\downarrow$}} &
        \thead{\textbf{BLEU}} &
        \thead{\textbf{ROUGE-1} \\ \textbf{F1}} & 
        \thead{\textbf{METEOR}} &
        \thead{\textbf{BERT} \\ \textbf{Score}} \\
        \midrule
        \multirow{6}{*}{\rotatebox[origin=c]{90}{\textsc{Beauty}}}
         & 1 & None & 0.67 & 0.61 & 0.78 & 0.30 & 0.69 & \textbf{1.24} & \textbf{1.68} & 0.241 & \textbf{0.510} & \textbf{0.339} & 0.667 \\
          & 8 & None & \textbf{0.68} & \textbf{0.64} & \textbf{0.79} & \textbf{0.31} & \textbf{0.70} & 1.25 & 1.71 & \textbf{0.248} & 0.509 & 0.333 & \textbf{0.671} \\
          & 8 & 5-class & 0.54 & 0.61 & 0.63 & 0.28 & 0.60 & 1.32 & 1.74 & \textbf{0.248} & \textbf{0.510} & 0.329 & 0.670 \\
          & 8 & Binary & 0.53 & 0.59 & 0.64 & 0.29 & 0.60 & 1.40 & 1.88 & 0.246 & 0.508 & 0.335 & 0.669 \\
          & 8 & 1-off & 0.61 & 0.61 & 0.71 & 0.30 & 0.63 & 1.28 & 1.75 & 0.247 & 0.336 & \textbf{0.510} & \textbf{0.671} \\
        \cmidrule(lr){2-14}
        & \multicolumn{2}{c}{PaLM 2-M Zero-shot} & 0.56 & 0.62 & - & 0.37 & - & 1.14 & 1.60 & 0.236 & 0.503 & 0.339 & 0.665 \\
        \cmidrule(lr){1-14}
        \multirow{6}{*}{\rotatebox[origin=c]{90}{\textsc{Movies/TV}}}
         & 1 & None & \textbf{0.65} & 0.61 & 0.72 & 0.34 & \textbf{0.67} & 1.17 & 1.64 & 0.165 & 0.449 & 0.286 & \textbf{0.643} \\
          & 8 & None & 0.63 & 0.58 & 0.72 & 0.35 & \textbf{0.67} & 1.23 & 1.75 & 0.171 & 0.446 & 0.285 & 0.642 \\
          & 8 & 5-class & 0.59 & 0.63 & 0.69 & 0.32 & 0.63 & 1.17 & \textbf{1.61} & 0.176 & 0.449 & \textbf{0.291} & 0.642 \\
          & 8 & Binary & 0.60 & \textbf{0.64} & 0.71 & 0.33 & 0.66 & 1.28 & 1.78 & 0.175 & 0.443 & 0.288 & 0.641 \\
          & 8 & 1-off & 0.62 & 0.63 & \textbf{0.74} & \textbf{0.36} & \textbf{0.67} & \textbf{1.16} & 1.64 & \textbf{0.180} & \textbf{0.451} & \textbf{0.291} & \textbf{0.643} \\
        \cmidrule(lr){2-14}
         & \multicolumn{2}{c}{PaLM 2-M Zero-shot} & 0.62 & 0.66 & - & 0.40 & - & 1.06 & 1.53 & 0.194 & 0.465 & 0.296 & 0.647 \\
        \bottomrule
    \end{tabulary}}
\end{table*}

\subsection{Fine-tuning with Reasoning Data}
We utilize Flan-T5 \cite{flan-t5} models as they are readily available to conduct fine-tuning experiments. Although these models are all trained on a variety of data and tasks, it should be noted that PaLM-2-M reports significantly higher quality than Flan-T5 on common benchmarks, including the Massive Multitask Language Understanding (MMLU) \cite{hendrycks2021measuring}. Unless specified, we employ the Flan-T5 XL model (3B parameters) to output reasoning followed by a final rating prediction. We fine-tune the models for 100,000 steps with a batch size of 64, a dropout rate of 0.25, and a learning rate of 1e-4. We report the evaluation metrics on the test set at the end of training. In Table \ref{tab:fine-tune-model-size} we compare two fine-tuned XL models: one that outputs reasoning and rating, and another that outputs rating only. For our first experiment, we use only 1 reasoning sample for each user-item pair, \textit{i.e.} $m=1$ for Eq. \eqref{eq:reason-gen}, and refrain from further filtering to maintain consistent training sample numbers for a fair comparison. We collect model probabilities by extracting and normalizing the logits corresponding to the 5 rating class tokens, allowing us to compute ROC-AUC metrics. The results from both domains consistently indicate that fine-tuning models to engage in reasoning prior to making predictions leads to improved performance across all metrics. 

In addition to comparing the XL models with reasoning outputs against those without, we fine-tune various Flan-T5 model sizes: Small (80M params), Base (250M), Large (780M), and XL (3B). All models are trained with one reasoning sample per example without any filtering. Table \ref{tab:fine-tune-model-size} illustrates a clear trend indicating that larger models perform better on the rating task. This observation aligns with the general understanding that larger models typically possess greater knowledge capacity, leading to enhanced performance on downstream tasks after fine-tuning. Moreover, larger models also tend to have better reasoning quality, which we discuss further in Section \ref{sec:eval-reasoning-quality}.

We also include results on the Flan-T5 XL model without any fine-tuning for reference, showing that fine-tuning is absolutely necessary to improve results for this model family. For this non fine-tuned model, we tried both with and without an additional reasoning output. When asked to output reasoning, the model was unable to follow instructions and did not output a final rating in almost all cases. Therefore, we only report no fine-tuning results without reasoning. Additionally, although we see drastic improvements for Flan-T5, our best Flan-T5 result still underperforms our best PaLM 2-M result without fine-tuning. This difference is likely attributable to the enhanced capabilities of PaLM 2-M relative to Flan-T5.

\vspace{-.2cm}
\paragraph{Training with multiple reasoning paths.}
In this experiment, we increase the number of reasoning samples $m$ to 8 for each user-item pair, as defined in Eq. \eqref{eq:reason-gen}, providing diverse reasoning paths for each user-item relation. We also apply different filtering methods based on comparing the zero-shot LLM rating predictions $\hat{\mathbf{r}}^m_{u, i}$ to the ground truth ratings $\mathbf{r}_{u, i}$. Table \ref{tab:fine-tune-more-samples-and-filtering} presents the results of training with multiple reasoning paths and with different filtering methods:
\begin{itemize}[topsep=0pt,itemsep=0ex,partopsep=1ex,parsep=0pt]
    \item \textbf{None}: No filter is applied. The model is trained with all reasoning samples $\hat{\mathbf{s}}^m_{u, i}$ per user-item pair.
    \item \textbf{5-class}: Reasoning samples are removed where $\hat{\mathbf{r}}^m_{u, i} \neq \mathbf{r}_{u, i}$ in terms of the 5-class rating.
    \item \textbf{Binary}: Reasoning samples are removed where the binary conversion of $(\hat{\mathbf{r}}^m_{u, i} > 3) \neq (\mathbf{r}_{u, i} > 3)$.
    \item \textbf{1-off}: Reasoning samples are removed where the absolute difference  $|\hat{\mathbf{r}}^m_{u, i} - \mathbf{r}_{u, i}| > 1$. 
\end{itemize}
In the \textsc{Beauty} domain, fine-tuning with 8 reasoning paths without any filtering slightly outperforms fine-tuning with only 1 reasoning path. Surprisingly, applying filtering methods significantly diminishes performance on the rating task. We hypothesize that filtering may remove a substantial portion of training samples, leading to poorer performance. This is particularly evident when the "5-class" filter, the most stringent filter, is applied. Conversely, in the \textsc{Movies/TV} domain, the best results are achieved with the "1-off" filtering method. We attribute this to the LLM's strong pre-trained knowledge in the \textsc{Movies/TV} domain, allowing it to tolerate the removal of examples where the reasoning does not align with the ground truth rating. In contrast, the \textsc{Beauty} domain may require more examples of user history and user-item relations for effective learning. Removing examples, even those with misaligned reasoning, may inadvertently reduce domain information, resulting in diminished performance. Further analysis of these effects will be conducted in future work.

\vspace{-.2cm}
\section{Reasoning Evaluation}
\label{sec:reasoning-eval}

\subsection{Human Judgment Alignment Analysis}
As proposed in Sec. \ref{sec:human-judgment-study}, we design a human judgement alignment study to evaluate the effectiveness of Rec-SAVER . We presented a total of 100 samples to human raters, with 50 examples from the \textsc{Beauty} domain and 50 examples from the \textsc{Movies/TV} domain. Each rating category was evenly represented. Each sample was annotated by 3 different annotators, resulting in a total of 300 annotated data points. Table \ref{tab:iaa-scores} presents the inter-annotator agreement of weighted Cohen $\kappa$ \cite{cohen} and the average Pearson correlation (Avg. $\rho$) among the human annotated scores. The achieved statistical significance of the average correlation between 3 human annotators across all 3 measurements, as indicated by the $p$-values, signifies the level of consensus among annotators. 

We evaluated four commonly used natural language generation (NLG) metrics: BLEU \cite{papineni2002bleu}, ROUGE-1 \cite{lin2004rouge}, METEOR \cite{banerjee2005meteor}, and BERTScore \cite{zhang2019bertscore}. BLEU and ROUGE-1 measure syntactic similarity by computing the exact $n$-gram overlap between the generated output and the reference texts. On the other hand, METEOR and BERTScore consider semantic similarity, providing a more comprehensive evaluation by incorporating contextual information. Table \ref{tab:correlation-nlg} reveals a consistently positive correlation between coherence and all NLG metrics, suggesting that our proposed evaluation method using LLM-generated references aligns well with coherence. However, insightfulness exhibits no correlation with BLEU and a low correlation with ROUGE-1 F1, while demonstrating a slightly positive correlation with METEOR and BERTScore. Unlike ``coherence'' and ''faithfulness'', ``insighfulness'' is an exploratory metric aimed at understanding how LLM \textit{surprise} human raters. It is anticipated that syntactic metrics such as BLEU and ROUGE-1 F1 may not correlate strongly with insightfulness. For instance, a response could be considered insightful even if it lacks significant $n$-gram overlap with the references. Although the semantic metrics METEOR and BERTScore show better correlation, they still do not align as closely with insightfulness as they do with coherence.

\begin{table}[!h]
    \centering
    \captionsetup{font=footnotesize}
    \footnotesize
    \renewcommand{\arraystretch}{0.95}
    \setlength{\tabcolsep}{3pt}
    \caption{\label{tab:iaa-scores} Inter-annotator agreement (IAA) analysis on the human annotated scores.}
    \begin{tabulary}{.5\textwidth}{LCCCR}
        \toprule
        & \textbf{Mean} & \textbf{Cohen $\kappa$} & \textbf{Avg. $\rho$} & \textbf{$p$-value} \\
        \midrule
        Coherence & 3.72 & 0.37 & 0.37 & 1e-10 \\
        Faithfulness & 0.63 & 0.63 & 0.63 & 1e-12 \\
        Insightfulness & 2.80 & 0.33 & 0.34 & 6e-4 \\
        \bottomrule
    \end{tabulary}
\end{table}

\begin{table}[!h]
    \centering
    \captionsetup{font=footnotesize}
    \footnotesize
    \renewcommand{\arraystretch}{0.95}
    \setlength{\tabcolsep}{3pt}
    \caption{\label{tab:correlation-nlg} Correlation between coherence, insightfulness, and automatic NLG metrics. The annotated scores are averaged across the annotators for each sample.}
    \begin{tabulary}{.5\textwidth}{LCC}
        \toprule
         & \textbf{Coherence} & \textbf{Insightfulness} \\
        \midrule
        BLEU & 0.36 & 0.02 \\
        ROUGE-1 F1 & 0.40 & 0.10 \\
        METEOR & 0.40 & 0.25 \\
        BERTScore & 0.36 & 0.20 \\
        \bottomrule
    \end{tabulary}
\end{table}

\vspace{-.2cm}
\paragraph{Two-sample T-test.}
We conducted a two-sample t-test to compare faithful and unfaithful reasoning. Faithful reasoning refers to outputs without factual or logical errors, while unfaithful reasoning refers to outputs containing one or more such errors. Table \ref{tab:t-test} shows that when errors are present, the reasoning responses are less coherent and insightful. Additionally, the average scores for all automatic NLG metrics are higher for faithful reasoning compared to unfaithful reasoning. Syntactic metrics (BLEU and ROUGE-1 F1) exhibit a more pronounced difference than semantic metrics (METEOR and BERTScore), as indicated by the lower $p$-values. This discrepancy arises because small differences in $n$-grams can lead to factual errors. For example, changing \textit{``the user purchased \textbf{4} products''} to \textit{``the user purchased \textbf{5} products''} can render a sentence unfaithful. Such small discrepancies are not as easily detected by the semantic metrics.

\begin{table}[!h]
    \centering
    \captionsetup{font=footnotesize}
    \renewcommand{\arraystretch}{0.95}
    \setlength{\tabcolsep}{5pt}
    \footnotesize
    \caption{\label{tab:t-test} Two-sample t-test comparing the average of human annotated scores and NLG scores between faithful and unfaithful reasoning.}
    \resizebox{.4\textwidth}{!}{%
    \begin{tabulary}{.75\textwidth}{lCCR}
        \toprule
         & \textbf{Faithful} & \textbf{Unfailthful} & \textbf{$p$-value} \\
        \midrule
         Coherence & \textbf{4.01} & 3.22 & \textbf{2e-8} \\
         Insightfulness & \textbf{3.11} & 2.23 & \textbf{6e-9} \\
         \cmidrule(lr){1-4}
         BLEU & \textbf{0.21} & 0.16 & \textbf{2e-3} \\
         ROUGE-1 F1 & \textbf{0.49} & 0.46 & \textbf{5e-3} \\
         METEOR & \textbf{0.31} & 0.30 & 0.36 \\
         BERTScore & \textbf{0.65} & 0.63 & \textbf{0.02} \\
        \bottomrule
    \end{tabulary}}
\end{table}

\vspace{-.3cm}
\paragraph{Effectiveness of Self-verification.}
To validate the effectiveness of the self-verification step in Rec-SAVER, we compared the metrics computed with and without the reference self-verification step. Table \ref{tab:self-verifcation} shows that metrics computed from self-verified references show a stronger correlation with the coherence score compared to those without self-verification, indicating that self-verification contributes to increasing the credibility of LLM-generated references. Combining all results of this section, we observe a strong alignment between our proposed Rec-SAVER reasoning evaluation method and human judgments regarding quality.

\begin{table}[!h]
    \centering
    \captionsetup{font=footnotesize}
    \footnotesize
    \renewcommand{\arraystretch}{0.95}
    \setlength{\tabcolsep}{7pt}
    \caption{\label{tab:self-verifcation} Correlation between coherence and NLG metrics with and without using self-verified references.}
    \begin{tabulary}{.5\textwidth}{LCC}
        \toprule
        \textbf{Self-verification} & \textbf{Yes} & \textbf{No} \\
        \midrule
        BLEU & \textbf{0.36} & 0.33 \\
        ROUGE-1 F1 & \textbf{0.40} & 0.35 \\
        METEOR & \textbf{0.40} & 0.37 \\
        BERTScore & \textbf{0.36} & 0.28 \\
        \bottomrule
    \end{tabulary}
\end{table}

\subsection{Analysis of Reasoning Quality} 
\label{sec:eval-reasoning-quality}

\begin{table*}[!h]
    \centering
    \captionsetup{font=footnotesize}
    \scriptsize
    \renewcommand{\arraystretch}{1.05}
    \caption{\label{tab:zero-shot-examples} Example reasons generated by PaLM 2-M in a zero-shot setting.}
    \begin{tabulary}{\textwidth}{lL}
        \toprule
        \textsc{Beauty} & The user has purchased 4 beauty products in the past, all of which are highly rated. The new product is also a beauty product, so it is likely that the user will be interested in it. The new product is a concealer, which is a type of makeup that is used to cover blemishes and imperfections. The user has not purchased any makeup products in the past, so it is possible that they are not familiar with concealers. However, the description of the new product states that is is ``natural coverage'' and ``non-nonsense'', which suggests that is is a good choice for beginners. Overall it is likely that the user will rate the new product positively. \\
        \cmidrule{1-2}
        \textsc{Movies/TV} & The user has given high ratings to movies that are inspiring and thought-provoking, and/or a history lesson. The new video is a historical movie about a group of black flyers who broke the color barrier in the U.S. Air Force during World War II. It is likely that the user will find this movie inspiring and thought-provoking, and therefore rate it highly. \\
        \bottomrule
    \end{tabulary}
\end{table*}

\begin{table}[!h]
    \centering
    \captionsetup{font=footnotesize}
    \footnotesize
    \renewcommand{\arraystretch}{1.10}
    \setlength{\tabcolsep}{5pt}
    \caption{\label{tab:reason-quality-correctness} Reasoning quality associated with correct and incorrect rating predictions for PaLM 2-M zero-shot and Flan-T5 XL fine-tuned (1 sample per example, no filtering) models.}
    \resizebox{.48\textwidth}{!}{%
    \begin{tabulary}{.75\textwidth}{LCCCCCC}
        \toprule
        &
        \thead{\textbf{Model}} &
        \thead{\textbf{Correct} \\ \textbf{Prediction}} & 
        \thead{\textbf{BLEU}} &
        \thead{\textbf{ROUGE-1} \\ \textbf{F1}} & 
        \thead{\textbf{METEOR}} &
        \thead{\textbf{BERT} \\ \textbf{Score}} \\
        \midrule
        \multirow{4}{*}{\rotatebox[origin=c]{90}{\textsc{Beauty}}}
         & PaLM 2-M  & Yes & \textbf{0.260} & \textbf{0.522} & \textbf{0.342} & \textbf{0.666} \\
         & PaLM 2-M  & No & 0.221 & 0.491 & 0.336 & 0.665 \\
         \cmidrule(lr){2-7}
         & Flan-T5 XL & Yes & \textbf{0.254} & \textbf{0.515} & \textbf{0.342} & \textbf{0.667} \\ 
         & Flan-T5 XL & No & 0.235 & 0.508 & 0.338 & 0.666 \\ 
        \cmidrule(lr){1-7}
        \multirow{4}{*}{\rotatebox[origin=c]{90}{\textsc{Movies/TV}}}
         & PaLM 2-M & Yes & \textbf{0.204} & \textbf{0.480} & \textbf{0.306} & \textbf{0.648} \\ 
         & PaLM 2-M & No & 0.187 & 0.455 & 0.290 & 0.647 \\
         \cmidrule(lr){2-7}
         & Flan-T5 XL & Yes & \textbf{0.177} & \textbf{0.457} & \textbf{0.292} & \textbf{0.644} \\ 
         & Flan-T5 XL & No & 0.159 & 0.444 & 0.283 & 0.642 \\ 
        \bottomrule
    \end{tabulary}}
\end{table}

Having established Rec-SAVER as a method for generating references and evaluating reasoning, we can leverage it to further analyze model reasoning outputs. We display example reasoning outputs in Table \ref{tab:zero-shot-examples}. Our first investigation focuses on the question: ``\textit{Are correct rating predictions generally associated with higher-quality reasons?}'' Table \ref{tab:reason-quality-correctness} demonstrates that reasoning metrics indeed improve for examples with correct predictions, both for zero-shot and fine-tuned models. Next, we analyze the reasoning quality across different methods discussed throughout the paper. Focusing on zero-shot models, we observe that in the \textsc{Movies/TV} domain, the NLG metrics decrease when we remove input information (Table \ref{tab:zero-shot-ablation-all}). However, for \textsc{Beauty}, when we remove some input information like user reviews, some of the NLG metrics increase. This suggests that the LLM may encounter challenges synthesizing all of the information in \textsc{Beauty} but possesses a better overall knowledge of \textsc{Movies/TV}, allowing it to generate better reasons when provided with more information available.

Table \ref{tab:fine-tune-model-size} illustrates that beyond the rating prediction metrics, the NLG metrics also improve as we increase the fine-tuned model size. This suggests that in addition to producing better rating predictions, the models also generate better reasoning responses. In the \textsc{Movies/TV} domain, the ``1-off'' filtering method appears to yield the best rating metrics (Table \ref{tab:fine-tune-more-samples-and-filtering}), although a few other methods are comparable. However, when considering the NLG metrics, we observe more data showing that the ``1-off'' filtering method has an advantage over the other methods.

The comparison between the fine-tuned models and the zero-shot model for \textsc{Movies/TV} reveals that the zero-shot model outperforms the fine-tuned model in both rating and NLG metrics. This outcome suggests that the PaLM 2-M zero-shot model likely possesses superior pre-trained knowledge in this domain, which cannot be fully distilled into the Flan-T5 fine-tuned model. For example, the reasoning text used in the fine-tuning training data may include mentions of certain actors or directors in movies. However, this data might not cover a wide enough range of examples to provide the model a comprehensive understanding of the entire domain. Consequently, the final fine-tuned model may still exhibit information gaps in this domain compared to the zero-shot model.

\section{Related Work}

\paragraph{LLM for RecSys.}
Recent advancements in the application of LLMs to RecSys have yielded diverse approaches. Typically, these approaches follow \textit{pre-training}, \textit{fine-tuning}, and \textit{prompting} paradigms. In the context of recommendation tasks, fine-tuning LLMs is essential to acquire domain-specific knowledge. This fine-tuning process involves training the pre-trained model using task-specific recommendation datasets containing user-item interaction behaviors such as purchase, ratings, or click, and additional contextual information about users and items (\textit{e.g.}, social relations or item descriptions) \cite{friedman2023leveraging, cui2022m6, liu2023llmrec}. Beyond the pre-training and fine-tuning paradigm, prompting has emerged as a recent paradigm to tailor LLMs for specific downstream tasks, employing task-specific prompts \cite{wang2022towards, ferraretto2023exaranker, gao2023chat}, along with prompting techniques like in-context learning \cite{gao2020making} and chain-of-thought \cite{wei2022chain}. More recently, there has been exploration into instruction tuning, a hybrid approach combining the pre-training and fine-tuning paradigm with prompting. This involves fine-tuning LLMs across multiple recommendation tasks using instruction-based prompts, enhancing the zero-shot performance of LLMs on previously unseen RecSys tasks \cite{geng2022recommendation, zhang2023recommendation, kang2023llms}.

\vspace{-.2cm}
\paragraph{LLM reasoning.}
Recent studies have suggested that the ability to reason may emerge in language models at a certain scale \cite{cobbe2021training, wei2022emergent}. These models, when provided with a few examples of ``chain of thought'', which represent intermediate natural language reasoning steps, demonstrate the capability to generate explicit rationales before producing final answers \cite{wei2022chain}. Advances in this direction include zero-shot CoT \cite{kojima2022large}, where the model is prompted with the phrase ``Let's think step by step'' to elicit reasoning without the inclusion of few-shot demonstrations. Various strategies have been proposed to enhance language model performance by prompting reasoning, such as multi-step reasoning \cite{dua-etal-2022-successive, zhou2023leasttomost}, tree-of-thoughts \cite{yao2023react}, iterative CoT prompting \cite{wang2022iteratively} and self-consistency \cite{wang2023selfconsistency}. Despite the impressive performance of LLMs on various reasoning tasks, the clarify of whether their predictions are based on true reasoning remains a challenge. This ambiguity arises because most existing evaluations focus on accuracy in end tasks rather than directly assessing the quality of the reasoning. Recent efforts have introduced metrics such as ROSCOE \cite{golovneva2022roscoe} and dataset such as PrOntoQA \cite{saparov2022language} for a more formal analysis of reasoning in LLMs.  However, the application of these metrics and benchmarks to a broader range of tasks is still an area of limited depth.

\section{Conclusion and Discussion}
We explore reasoning in the context of personalized recommender systems, showing that adding reasoning steps can improve LLM task performance. It is important to have rich user context and explicit feedback in order for the LLMs to reason adequately. Having good pre-trained domain knowledge is also useful. \textbf{RecSAVER}, our proposed method for analyzing reasoning quality, aligns well with human judgment on the coherence of reasoning outputs and can be used to further evaluate model quality beyond numerical task results.

\paragraph{Limitations.}
In this work we started with rating predictions in the Amazon review dataset for two categories, \textsc{Beauty} and \textsc{Movies/TV}. However, the extent of recommender systems is vast.  It is unclear to what extent our methods generalize more broadly to other categories such as music, video games, website articles, etc. Furthermore, more work is needed to explore these methods on other tasks, including candidate retrieval or ranking.

Now that we see evidence that reasoning is useful in RecSys, more work should be done to understand the extent and mechanisms behind this. Does the LLM actually reason in a manner that helps make a final decision similar to human thought? Or is there some other underlying procedure that yields these results, such as more overall computation or better attention? Future work looking at different prompting strategies and reasoning plans could help uncover more details in this area.

\paragraph{Ethical Considerations.}
In this study, biases may exist for reasoning results for different users, including users that speak different languages or users with different genders. The dataset we use focuses on users that speak English. Also, users from different genders may interact differently with certain domains or with products in those domains, leading to skewed distributions in the data. Broader experiments are needed to understand these potential biases further in the context of reasoning for recommender systems.

\section{Acknowledgements}
We would like to thank Jianmo Ni, Nikhil Mehta, Ramkumar Rajendran, and Lakshmi Chakrapani for their helpful discussions and support.

\clearpage
\bibliography{references}
\bibliographystyle{stylefiles/acl_natbib}

\clearpage
\appendix

\section{Reasoning Generation Prompt}
\label{app:reasoning-generation-prompt}

\begin{table}[!h]
    \captionsetup{font=footnotesize}
    \footnotesize
    \renewcommand{\arraystretch}{0.95}
    \caption{\label{tab:reason-gen-prompt} Reasoning generation prompt used in our zero-shot and fine-tuned experiments.}
    \resizebox{.48\textwidth}{!}{%
    \begin{tabulary}{.5\textwidth}{L}
    \toprule
    Here is information about a user and a new \texttt{\{product / video (movies and tv)\}} being recommended to the user. For the user, we have the user's past item  information history and the user's corresponding ratings. User ratings range from 1 to 5, where 1 is the lowest and 5 is the highest. For the new item being recommended, we have the item information. 
    \newline\newline
    \#\#\# Past User History: \#\#\# 
    \newline
    \texttt{\{Product / Video (Movies and TV)\}} Title: \texttt{\{\texttt{title}\}} \newline
    Brand: \texttt{\{brand\}} \newline
    Categories: \texttt{\{categories\}} \newline
    Description: \texttt{\{description\}} \newline
    Item Price: \texttt{\{price\}} \newline
    User Rating: \texttt{\{userRating\}} \newline
    User Review: \texttt{\{reviewText\}} \newline
    . \newline
    . \newline
    . \newline
    
    \#\#\# New Item Information: \#\#\#
    \newline
    New \texttt{\{Product / Video (Movies and TV)\}} \newline
    \texttt{\{Product / Video (Movies and TV)\}} Title: \texttt{\{title\}} \newline
    Brand: \texttt{\{brand\}} \newline
    Categories: \texttt{\{categories\}} \newline
    Description: \texttt{\{description\}} \newline
    Item Price: \texttt{\{price\}} \newline
    
    \#\#\#\#\#\#
    \newline\newline
    Given the user's past \texttt{\{purchase / watch\}} history and the new item information, what information can you infer about the user's preferences and how they will rate the new \texttt{\{product / video (movies and tv)\}}
    ?
    \newline\newline
    Your reasoning explanation should be based on any commonalities in the user history items and inferred user tastes or preferences.
    \newline\newline
    After your reasoning, predict a numerical rating.
    \newline\newline
    Please follow the format below: \newline
    \#\#\# Reason \#\#\# \newline
    Write your reasoning explanation here. You can have line breaks.
    \newline\newline
    \#\#\# Rating \#\#\# \newline
    Give a single numerical rating, e.g. 1 \\
    \bottomrule
    \end{tabulary}}
\end{table}

\section{Additional Experimental Results}
\label{app:extra-stuff}

The weighted Cohen $\kappa$ is calculated as followed:
\[\kappa = 1- \frac{\sum_{i=1}^n \sum_{j=1}^n w_{ij} x_{ij}}{\sum_{i=1}^n \sum_{j=1}^n w_{ij} m_{ij}},\]
where $n=5$ is the number of rating scale, and $w, x$ and $m$ are elements in the weight, observed, and expected matrices. Here, we use a quadratic weight where $w_{ij} = \frac{(i-j)^2}{(k-1)^2}$ to amplify the difference between scores.

Table \ref{tab:additional-incorrect-correct-pred} provides a comparison of human-rated reasoning outputs between incorrect and correct predictions, as depicted in Figure \ref{fig:reasons-by-predict}. The table shows the mean scores for coherence, faithfulness, and insightfulness for reasoning outputs associated with incorrect and correct predictions. We observe that reasoning outputs corresponding to correct predictions receive higher scores across all three dimensions, indicating a higher reasoning quality when the prediction is correct.

\begin{figure}[!h]
    \centering
    \captionsetup{font=footnotesize}
    \includegraphics[width=.5\textwidth]{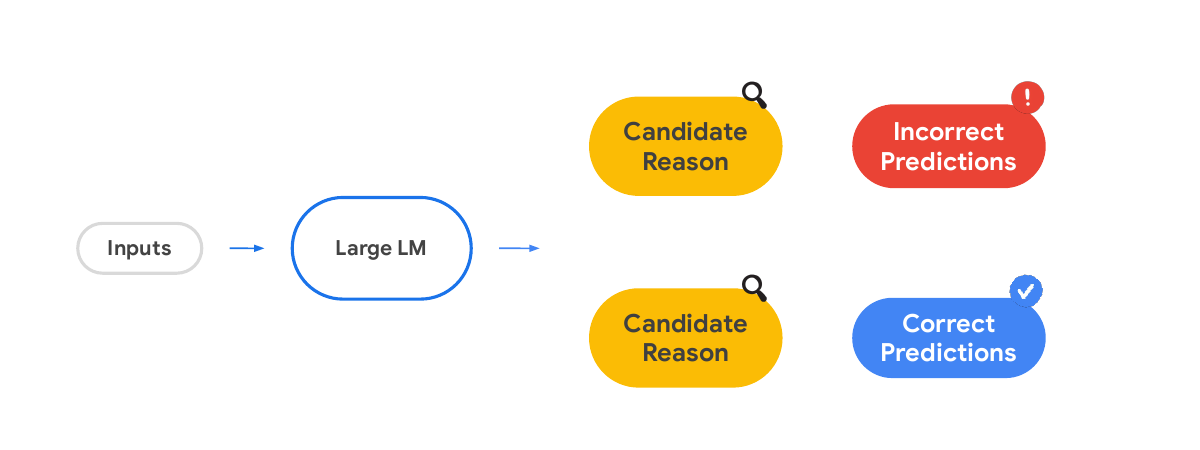}
    \caption{Outputs reasons categorized based on the correctness of rating predictions.}
    \label{fig:reasons-by-predict}
\end{figure}

\begin{table}[!h]
    \centering
    \captionsetup{font=footnotesize}
    \footnotesize
    \renewcommand{\arraystretch}{.95}
    \caption{\label{tab:additional-incorrect-correct-pred} Comparison of human-rated reasoning outputs between incorrect and correct predictions. Higher scores indicate higher reasoning quality.}
    \resizebox{.48\textwidth}{!}{%
    \begin{tabulary}{\textwidth}{lRR}
        \toprule
         & \textbf{Incorrect Prediction} & \textbf{Correct Prediction} \\
        \midrule
        Coherence & 3.59 & 3.91 \\
        Faithfulness & 0.61 & 0.67 \\
        Insightfulness & 2.71 & 2.93 \\
        \cmidrule{1-3}
        \textbf{Types of Errors} & & \\
        \cmidrule{1-3}
        Incorrect Product Statistics & 28\% & 21\% \\
        Incorrect Product Information & 14\% & 16\% \\
        Arithmetic Errors & 5\% & 3\% \\
        Others & 2\% & 1\% \\ 
        \bottomrule
    \end{tabulary}}
\end{table}

\begin{table}[!h]
    \centering
    \footnotesize
    \renewcommand{\arraystretch}{1.05}
    \caption{\label{tab:fine-tuned-example} Example output generated by the fine-tuned FLAN-T5 XL model.}
    \begin{tabulary}{.48\textwidth}{L}
        \toprule
        \#\#\# Reason \#\#\# \newline
        The user has a history of watching action movies, especially those with a sci-fi or fantasy element. The new video is an action with a Batman theme, so it is likely to appeal to the user. \newline\newline
        \#\#\# Rating \#\#\# \newline
        5
        \\
        \bottomrule
    \end{tabulary}
\end{table}

\begin{table}[!h]
    \centering
    \captionsetup{font=footnotesize}
    \footnotesize
    \renewcommand{\arraystretch}{1.05}
    \caption{\label{tab:nlg-deeper-dive} Comparing NLG metric statistics for a fine-tuned FLAN XL model and a zero-shot model in \textsc{Beauty}.}
    \resizebox{0.48\textwidth}{!}{%
    \begin{tabulary}{\textwidth}{LLCCCC}
        \toprule
        &  & 
        \thead{\textbf{ROUGE-1} \\ \textbf{F1}} & 
        \thead{\textbf{BLEU}} &
        \thead{\textbf{METEOR}} &
        \thead{\textbf{BERTScore}} \\
        \midrule
        \multirow{4}{*}{Fine-Tuned}
         & Mean & 0.509 & 0.248 & 0.333 & 0.671 \\
         & Min & 0.256 & 0.028 & 0.163 & 0.524 \\
         & Max & 0.804 & 0.771 & 0.575 & 0.816 \\
         & Range & 0.548 & 0.743 & 0.412 & 0.292 \\
         & Std-Dev & 0.087 & 0.131 & 0.073 & 0.046 \\
        \cmidrule(lr){1-6}
        \multirow{4}{*}{Zero-Shot}
         & Mean & 0.506 & 0.245 & 0.332 & 0.665 \\
         & Min & 0.088 & 0.010 & 0.110 & 0.480 \\
         & Max & 0.852 & 0.740 & 0.772 & 0.838 \\
         & Range & 0.765 & 0.730 & 0.662 & 0.359 \\
         & Std-Dev & 0.090 & 0.132 & 0.076 & 0.047 \\
        \bottomrule
    \end{tabulary}}
\end{table}

\end{document}